# Increasing countries' financial resilience through global catastrophe risk pooling

Alessio Ciullo [1]*, Eric Strobl [2], Simona Meiler [1], Olivia Martius [3], David Bresch [1,4]

[1] Institute for Environmental Decisions, ETH Zurich, Zurich, Switzerland
[2] Department of Economics, University of Bern, Switzerland
[3] Institute of Geography, University of Bern, Switzerland
[4] Swiss Federal Office of Meteorology and Climatology MeteoSwiss, Zurich, Switzerland

* Corresponding author, alessio.ciullo@usys.ethz.ch

## Abstract

Extreme weather events can have severe impacts on national economies, leading the recovery of low- to middle-income countries to become reliant on foreign financial aid. Foreign aid, however, is slow and uncertain. Therefore, the Sendai Framework and the Paris Agreement advocate for more resilient financial instruments like sovereign catastrophe risk pools. Existing pools, however, might not fully exploit financial resilience potentials because they were not designed with the goal of maximizing risk diversification and they pool risk only regionally. To address this, we introduce a method that forms pools maximizing risk diversification and which selects countries with low bilateral correlations or low shares in the pool's risk. We apply the method to explore the benefits of global pooling with respect to regional pooling. We find that global pooling increases risk diversification, it lowers countries' shares in the pool's risk and it increases the number of countries profiting from risk pooling.

## Introduction

Extreme weather events like tropical cyclones, floods, and heavy precipitation can have severe impacts on economies leading to a short-term deterioration of several macro-economic variables. In the Caribbean region, for example, an average hurricane strike was found to cause a loss in the growth of annual aggregate of about 0.84 %[1], a local income growth loss of 1.5 %[2], a cumulative expected loss in total tax revenue of 5.3 %[3], a multifold increase in monthly average inflation due to shortages of goods[4] and an appreciation of real exchange[5].

These deteriorated macro-economic scenarios are likely to require increases in governments spending[6] via short-term deficit financing, which in turn leads to debt increase[3]. For countries facing pre-existing debt sustainability issues this may be very costly[7] and, therefore, their recovery often relies on financial aid from international donors acting as insurers of last resort. Although foreign financial aid can help mitigate the effect of natural disasters on economic growth[8], it is also generally considered to be a slow and uncertain *ex-post* financial instrument[9]. Foreign financial aid may take months to materialize and it is impossible to *a priori* assess what amount, if any, will be provided and under what conditions. Indeed only about 60 % of the humanitarian requests are covered and funds have not been equally allocated between emergencies[10,11]. In contrast, *ex-ante* financial instruments, e.g., insurance, provide faster and more predictable flows of funding in the aftermath of disasters and allow governments to spread costs over time at a predictable rate[10]. Furthermore, *ex-ante* financial instruments complement non-financial disaster risk management strategies as they may foster investments in risk reduction and increase preparedness and adaptation[11].

Several international high-level policy agendas advocate for the strengthening of financial resilience towards the impact of extreme natural hazards via *ex-ante* financial instruments[12]. For instance, the 2015 Sendai Framework for Disaster Risk Reduction promoted by the United Nations outlines four actions to prevent and reduce disaster risk. In this regard, the framework's third action relates to the importance of *ex-ante* investments for reducing disaster risk and increasing resilience via insurance and risk-sharing mechanisms to reduce financial impacts on governments[13]. Also, Article 8 of the Paris Agreement reaffirmed the Warsaw International Mechanism for Loss and Damage and promoted risk insurance facilities and climate risk pooling as areas of cooperation and facilitation[14]. Following these calls, the *InsuResilience* Global Partnership was launched by the G20 and V20 Groups at COP23 in November 2017. *InsuResilience* identifies sovereign catastrophe risk pools, a financial mechanism where different countries pool their risk into a single portfolio, as being a promising *ex-ante* instrument especially for countries with low geographical (e.g., due to a limited size) or temporal (e.g., due to a limited borrowing capacity) risk spreading potential[9].

An effective risk pooling makes countries' share of the pool's risk lower than their individual risks[15] and, therefore, it lowers countries' technical premiums with respect to when they buy insurance separately. In particular, the technical premium is mainly determined by three factors: operational costs, cost of capital and annual expected loss[16]. Risk pooling reduces operational costs and the cost of capital. Operational costs are reduced because they are shared among all countries in the pool thus enabling economies of scale. A reduction in the cost of capital is achieved via an increased financial efficiency and it provides the largest premium reduction[11,16]. Financial efficiency is primarily reached via risk diversification, which stems from the fact that large losses will not be experienced by all countries simultaneously and therefore insuring the pooled risk requires much less capital than insuring all individual risks separately[10,17]. Financial efficiency is also increased via the establishment of joint reserves which allow retaining a larger risk share than what countries could individually retain, thus reducing the fraction of risk transferred to the reinsurance market. Furthermore, a reduction in the costs of reinsurance is achieved thanks to larger excess risk transactions to the reinsurance market.

Currently three sovereign catastrophe risk pools exist: the Caribbean Catastrophe Risk Insurance Facility (CCRIF), the African Risk Capacity (ARC), and the Pacific Catastrophe Risk Assessment and Financing Initiative (PCRAFI). CCRIF and PCRAFI cover tropical cyclone, excess rainfall and seismic risks, ARC covers mainly drought risk and, for few countries, also tropical cyclone and flood risk. Although these pools provide significant diversification benefits, their risk diversification might be limited for two reasons (World Bank, 2017). First, these pools were not designed with the main goal of maximizing risk diversification, but to rather serve the interest of their individual members. Second, the three pools are all regional and thus may not benefit from further potential risk diversification stemming from the inclusion of countries located outside their regional borders.

This paper addresses the issue of maximizing risk diversification and of expanding regional pools beyond their borders. It first introduces a method to find *optimal* risk pools, i.e., those with the highest achievable risk diversification reached with the least number of countries, and it then applies the method to assess and compare potential risk diversification benefits stemming from regional and global optimal pooling. We first identify the optimal regional pools for four geographical regions prone to tropical cyclones, and we assess to what extent global pooling might improve their risk diversification. We then focus on the two existing regional pools covering tropical cyclone risk, i.e., CCRIF and PCRAFI, and we assess their current risk diversification and the extent to which they might benefit from regional and global optimal pooling.

## Results

We identify four geographical regions prone to tropical cyclones: East Asia & Pacific (EAP), Central America & Caribbean (CAC), South Asia (SA) and South-East Africa (SEA) (see also Figure S1). The EAP region comprises 26 countries, the CAC region 38, the SEA region 16 and the SA region only 7. The four regions are identified following the World Bank's official regional classification and retaining only middle- to low-income countries facing tropical cyclone risk.

A 10000-years series of total annual tropical cyclone losses is reconstructed to assess risk diversification of sovereign catastrophe pools (*pools* for short in the aftermath) (see *Method*). The pools' risk diversification is assessed considering the 200-year event, which implies an $\alpha$ of 0.995 when calculating the Value-at-Risk, *VaR*, the Expected Shortfall, *ES,* and the Marginal Expected Shortfall, *MES* (see *Method*).

In the aftermath, when reporting correlations of losses between countries, these refer to the yearly total losses higher than the 200-year loss and they are calculated using the Pearson correlation coefficient. Countries are reported via their ISO 3166-1 alpha-3 codes, the reader is referred to Tables S1 – S4 to match countries' ISO codes with their official names.

### Regional Optimal Pools

Finding the optimal regional pools for each of the four regions requires carrying out the first optimization step introduced in *Method* for one pool at a time solving four single-objective optimization problems. Results are shown in Figure 1. The optimal pool in the CAC region has the highest diversification (0.75), followed by those in the EAP (0.66), SEA (0.5) and SA (0.33) regions (top panel in Figure 1). Risk diversification potentials are thus higher when more countries can potentially join the pool.

A pool's risk diversification and its composition depend on the countries' correlation structure (first column from the second to last row in Figure 1). The optimal pools are mostly made of uncorrelated or poorly correlated countries within a region. This follows from obvious risk diversification considerations, as highly correlated countries are likely to simultaneously experience losses and therefore decrease the pool's risk diversification. For example, in CAC, the region which exhibits the highest inter-regional correlations, countries like Anguilla (AIA), Saint-Barthélemy (BLM), Saint Martin (MAF) and Sint Maarten (SXM) have high reciprocal correlations ranging from 0.85 (AIA and BLM) to 0.95 (MAF and SXM, and MAF and BLM) and are left out from the optimal pool. The same applies to Saint Kitts and Nevis (KNA) and Montserrat (MSR), which have a correlation of 0.75. Similar considerations can be drawn for the other regions, where Viet Nam (VNM) and Cambodia (KHM) in EAP, Bhutan (BTN) and Bangladesh (BGD) in SA, Zimbabwe (ZWE) and South Africa (ZAF) or Somalia (SOM) and Ethiopia (ETH) in SEA exhibit the highest correlations within their region and are not part of the respective regional optimal pool. All these high correlations are explained by the geographical proximity of the countries involved.

Bilateral correlations among countries do not fully explain the pools' composition, as this also depends on the shares of countries' individual risk contributing to the pool's risk (see *Method*). These are shown in the second column from the second to last row in Figure 1. In CAC, for example, Barbados (BRB) and Saint Lucia (LCA) have a correlation of 0.54 and they are both part the optimal regional pool. Similarly, in EAP, Samoa (WSM) and American Samoa (ASM) are both part of the optimal regional pool and exhibit a correlation of 0.30. This occurs because the individual share of risks of these countries in the optimal pools is very low (0.12 for BRB and 0.15 for LCA, 0.06 for WSM, 0.03 for ASM) which implies that, although risks are correlated, only a small fraction contribute to the overall pool risk. On the contrary, countries highly correlated with other countries with a high individual concentration are left out. This is the case of Panama (PAN) which is not part of the optimal regional CAC pool. Panama has a correlation of 0.45 with Colombia (COL), which has the highest share of individual risk in the optimal CAC pool (0.5).

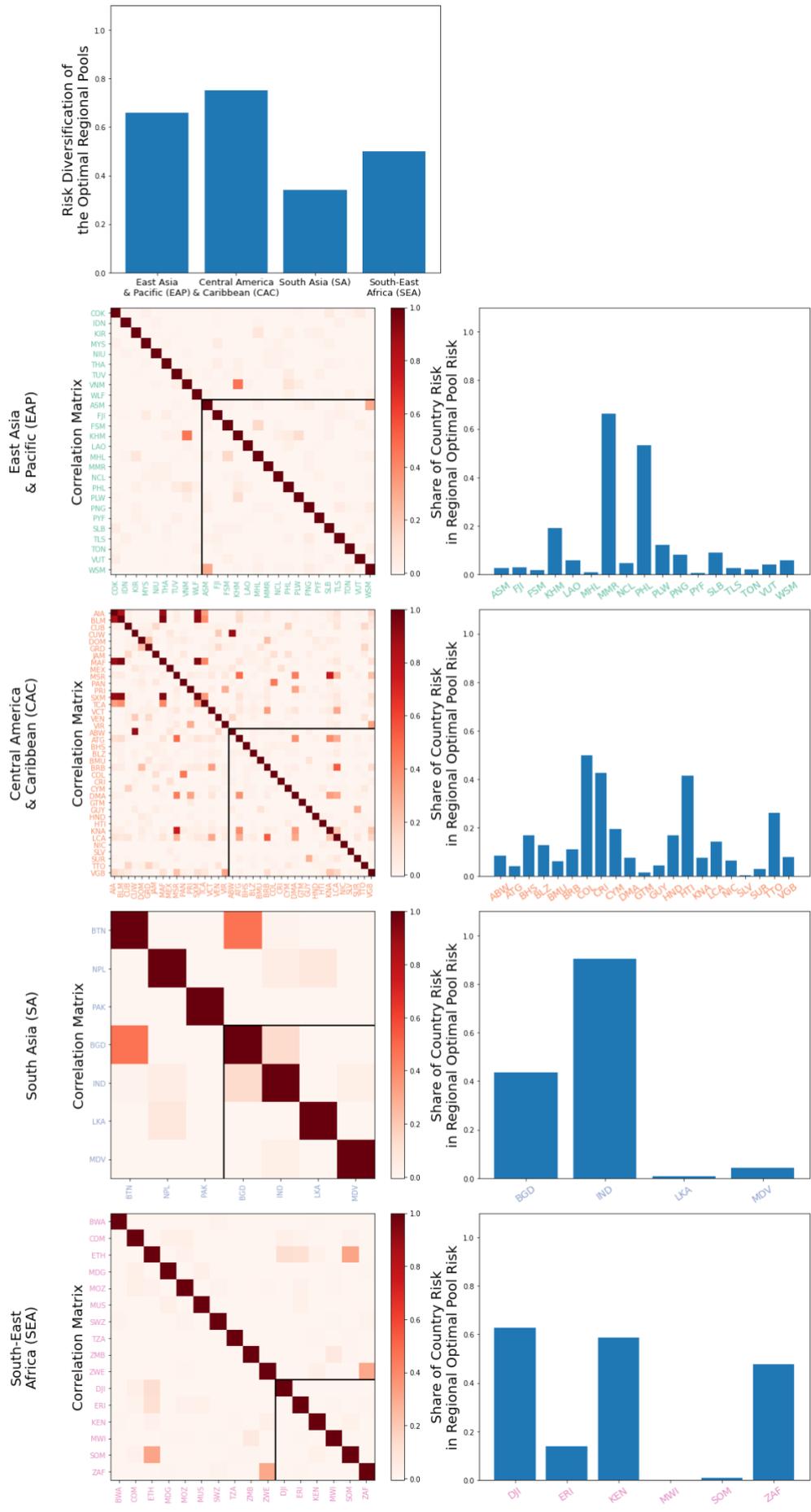

*Figure 1* Results for the optimal regional pools in the East Asia & Pacific (EAP), Central America & Caribbean (CAC), South Asia (SA) or South-East Africa (SEA) regions. The top bar plot shows risk diversification of the four regional optimal pools. Plots from the second to last row show correlation matrices and the share of countries' risk contributing to the pool's risk within each region. The correlation matrixes show the Pearson correlation coefficient for impacts with a return time of 200-y or higher for all countries in the region (full matrix) and countries within the optimal pool (sub-matrix within the black line). Share of countries' risk contributing to optimal regional pools' risks are shown as bar plots. Countries are reported via their ISO 3166-1 alpha-3 codes, and they are colored light green, orange, light blue or pink if they respectively belong to the EAP, CAC, SA or SEA region.

## Globally Diversified Regional Optimal Pools

After finding the optimal regional pools, we explore whether - and to what extent - possible global expansions of these pools increase their risk diversification. In doing so, the search for new countries that could join an optimal regional pool is global and no longer bounded within a given region. Any country not previously included in the optimal pool of its own region may join any - but only one - of the globally expanded regional optimal pools. Thus, it follows that optimal global pooling needs to be carried out simultaneously for the four regional pools via a four-objectives optimization (see *Method*). Results are shown in Figure 2.

The first row of the figure shows that many possible configurations of the four globally extended regional optimal pools exist. All these configurations increase risk diversification for all four pools, implying that global pooling leads to a strong Pareto improvement of the regional optimal pools. The extent of such increase differs across regions. Regions where optimal regional diversification was the lowest, i.e., SA and SEA, benefit most from global pooling. More precisely, the highest achievable diversification via global pooling doubles for SA (from 0.34 to 0.7) and reaches a 40 % increase for SEA (from 0.5 to 0.7). In EAP and CAC, where optimal regional diversification was already high, diversification increase is less prominent, and amounts to a maximum of about 15 % for EAP (from 0.66 to 0.75) and of about 6.5 % for CAC (from 0.75 to 0.8). The four pools therefore reach comparable maximum risk diversifications after global pooling.

The maximum risk diversification is, however, not achievable for all four pools together as trade-offs exist among the various optimal configurations of the four globally extended regional pools. The trade-off is particularly relevant for SA and SEA, as the highest diversification is reached when it is lowest for SEA. Such a trade-off is explained by the fact that there are some countries, i.e., Malaysia (MYS), Viet Nam (VNM), Cuba (CUB), Dominican Republic (DOM), Jamaica (JAM), Mexico (MEX), Panama (PAN) and Tanzania (TZA) that are part of the best globally extended regional pools of both regions.

The second to last rows of Figure 2 show individual countries' risk share contributing to the pool's risk and the composition of each pool. Overall, global pooling tends to decrease all countries' risk shares contributing to the pool's risk, and this happens because the pool's risk is redistributed elsewhere across the globe. Interestingly, global pooling also allows some regions, e.g., SEA and CAC, to pool countries within their own region which were not previously selected in the optimal regional pooling. This occurs because global pooling decreases the risk share of these countries and thus allows them to effectively join their own regional pool. It happens even within correlated countries, as it is the case for, e.g., Sint Maarten (SXM) and Turks and Caicos Islands (TCA), which are both part of the globally diversified CAC pool with a very low risk share (0.09 for SXM and 0.03 for TCA) despite a moderate bilateral correlation (0.35).

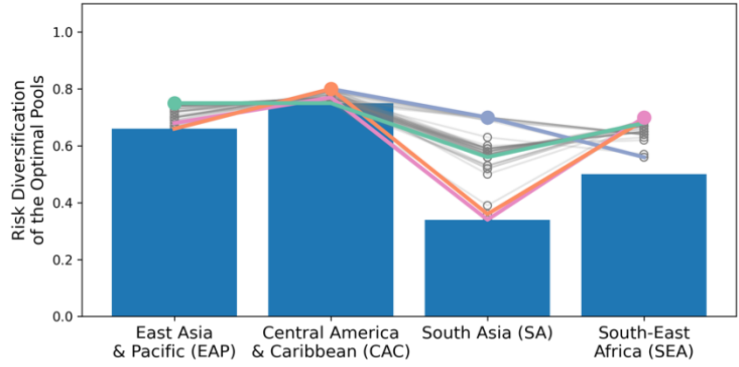

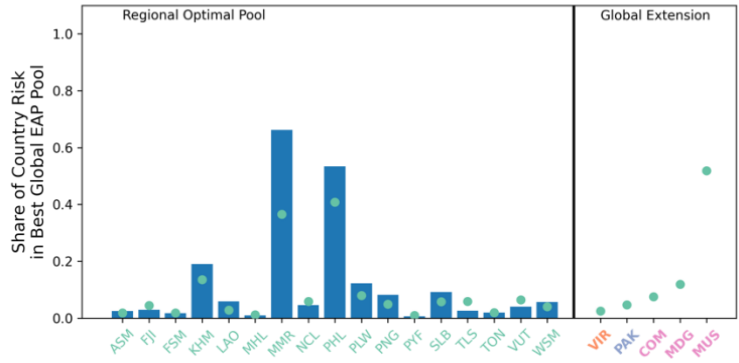

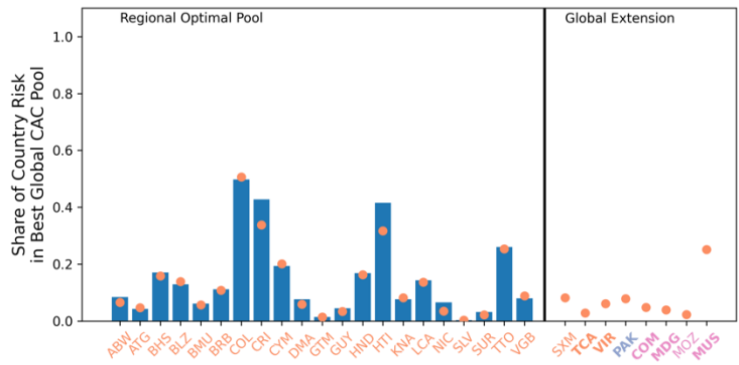

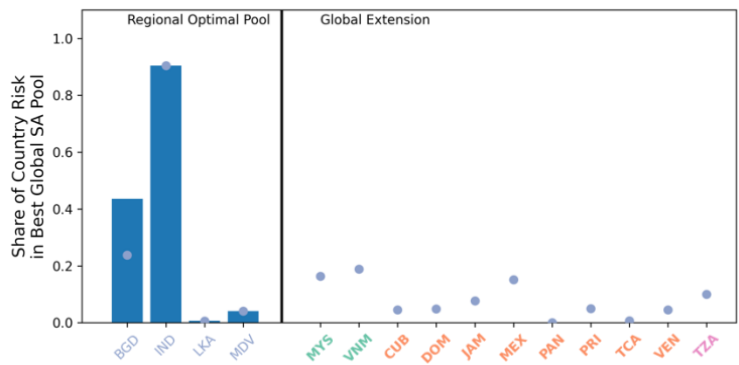

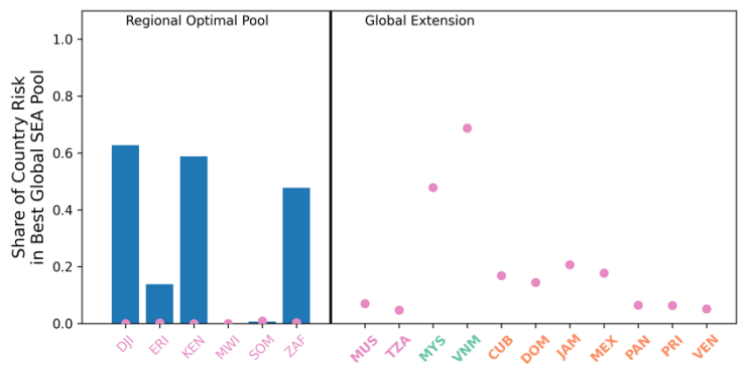

*Figure 1* Results for the globally diversified optimal regional pools for the East Asia & Pacific (EAP), Central America & Caribbean (CAC), South Asia (SA) or South-East Africa (SEA) regions. The top plot shows risk diversifications of the four regional optimal pools (bars) and the various configurations of the globally diversified regional optimal pools (continuous lines). For the latter, all configurations are reported in gray and the best configuration for each region is highlighted in light green, orange, light blue or pink if it refers to the EAP, CAC, SA or SEA region, respectively. The highest diversification for each region is indicated with a dot following the same coloring scheme. Plots from the second to last row show, for each region, the share of countries' risk contributing to the regional optimal pool's risk (bars) and the best globally diversified optimal regional pool's risk (dots). Countries are reported via their ISO 3166-1 alpha-3 codes following the aforementioned coloring scheme. ISO codes reported in bold are present in more than one best globally diversified optimal regional pool.

## Regional and Global Optimal Diversification of PCRAFI and CCRIF

After finding the optimal regional pools for the four regions and assessing the effect of optimal global pooling on risk diversification, we now focus on the two existing pools providing coverage for tropical cyclone risk: PCRAFI and CCRIF. We assess their current risk diversification and explore to what extent regional and global optimal expansions of these pools increase their risk diversification. Results are reported in Figure 3.

The top row of Figure 3 shows that an extension via optimal regional pooling (yellow crosses) leads to a diversification increase of 35 % for PCRAFI (from 0.49 to 0.66) and of about 40 % for CCRIF (from 0.48 to 0.67). In the case of PCRAFI, a diversification of 0.66 is the maximum reachable regional diversification since it equals the diversification of the optimal regional pool in the EAP region. For CCRIF, on the contrary, the achieved risk diversification via optimal regional pooling is about 89 % of the achievable maximum diversification in the CAC region. This implies that the initial design of CCRIF prevents exploiting the full diversification potential within its region.

In terms of individual countries' share of risk contributing to the pool's risk (first and second columns from first to last row in Figure 3), most countries in both PCRAFI and CCRIF have low shares already in the original pool (blue bars). There are few exceptions of countries with high shares, like Papua New Guinea (PNG) (almost 1.0) in PCRAFI or Jamaica (JAM) in CCRIF (also very high, 0.94). After regional pooling (yellow cross) Papua New Guinea substantially lowers its risk share to 0.09 while Jamaica lowers it only to 0.60. Jamaica is also the country with the largest modeled losses within CCRIF. This large concentration of CCRIF's risk on a single country explains why the pool cannot exploit the full diversification potential within the region.

The top row of Figure 3 also shows that there are three possible configurations of globally diversified PCRAFI and CCRIF. All these configurations have a higher diversification than the original pools (blue bars) and the regionally diversified original pools (yellow crosses). This confirms that global pooling leads to a Pareto improvement of regionally diversified pools. The highest possible diversification is higher in PCRAFI (0.81, i.e., a 65 % increase from its initial value) than in CCRIF (0.77, i.e., a 60 % increase from its initial value). Although a trade-off in increasing risk diversification for the two pools exists, this does not seem to be relevant since the difference in risk diversification for the three possible globally diversified CCRIF pools ranges within 2 percentage points (from 0.75 to 0.77). Thus, only one configuration is selected for further exploration, i.e., the one leading to the highest PCRAFI diversification (dotted line in purple).

For the selected configuration, the globally diversified PCRAFI gets a larger set of additional countries than the globally diversified CCRIF. Both PCRAFI and CCRIF pool many countries from their own region but PCRAFI, in addition, also pools many countries from CAC. A lower number of countries are pooled from SEA and SA. Papua New Guinea (PNG) and Jamaica (JAM), the two countries with the highest risk share in the original PCRAFI and CCRIF pools, substantially decrease their risk share after global pooling, as it was the case for regional pooling. Unlike regional pooling, however, global pooling does not increase risk shares in any other country in the region. This occurs because, in the globally diversified pools, countries with the highest risk shares belong to another region and are thus uncorrelated. In the globally diversified PCRAFI, the countries with the highest share are Colombia (COL) (0.59) and Costa Rica (CRI) (0.33) in the CAC region, and Mauritius (MUS) (0.31) in the SEA region. In the globally diversified CCRIF, the countries with the highest risk share are Malaysia (MYS) (0.42) and Viet Nam (VNM) (0.5) in the EAP region, and Bangladesh (BGD) (0.43) in the SA region.

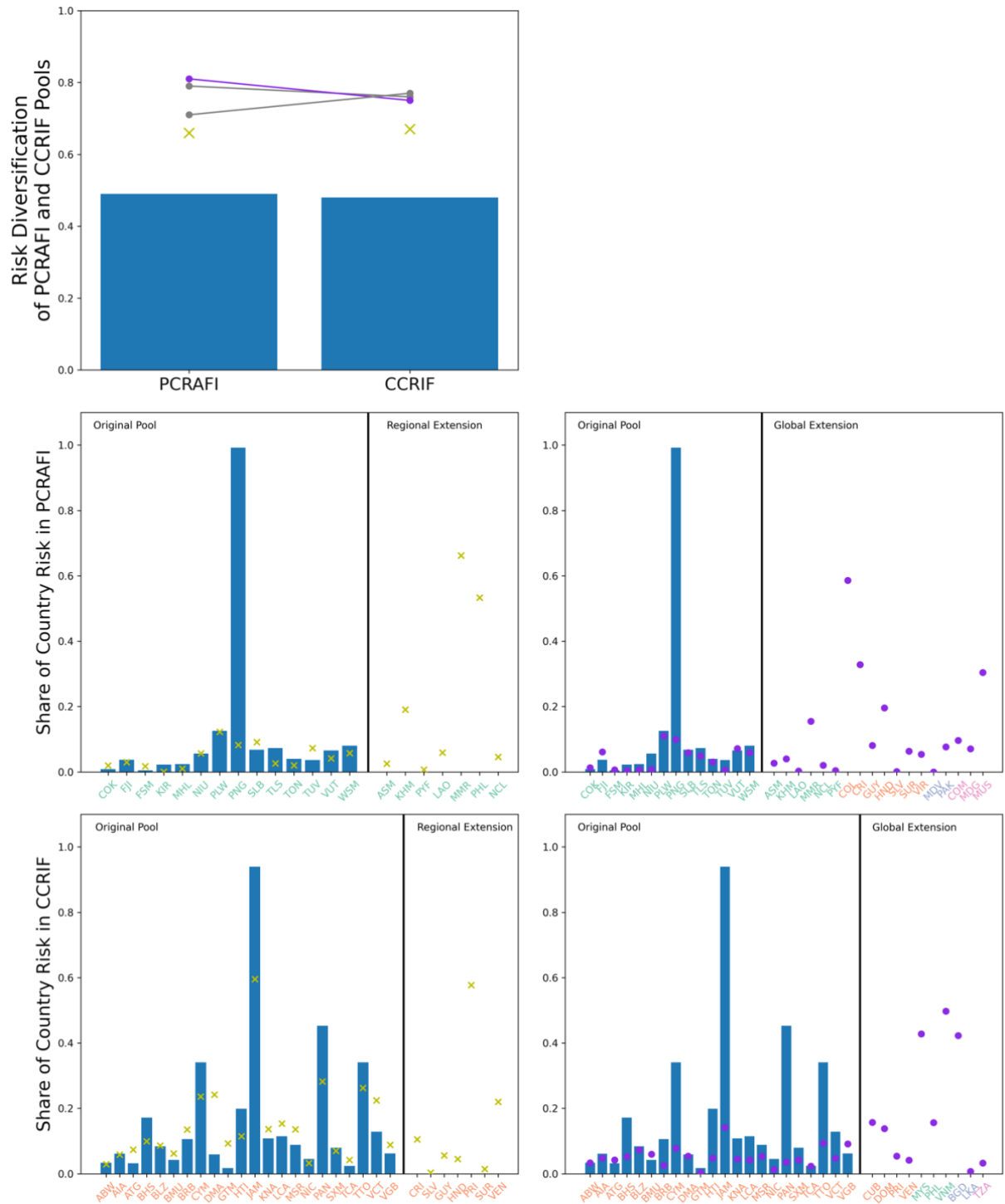

*Figure 2* Results of the regional and global optimal diversification for PCRAFI and CCRIF. The top plot shows risk diversifications of the original pools (bars), the regionally (yellow cross) and the globally (solid lines) diversified pools. Regarding the latter, all configurations are reported in gray, and the selected one is highlighted in purple. Plots in the second to the third rows show the shares of countries' risk contributing to the original PCRAFI's (second row) and CCRIF's (third row) risks and to their regionally (first column) and globally (second column) diversified pool's risks. Countries are reported via their ISO 3166-1 alpha-3 codes, and they are colored of light green, orange, light blue or pink if they respectively belong to the East Asia & Pacific (EAP), Central America & Caribbean (CAC), South Asia (SA) or South-East Africa (SEA) region.

# Discussion

Several international high-level policy agendas like the Sendai Framework and the Paris Agreement advocate for strengthening countries' financial resilience toward the impact of extreme natural hazards via *ex-ante* financial instruments. These instruments are considered resilient because they guarantee a predictable flow of funding in the aftermath of disasters and thus allow governments to spread costs over time at a predictable rate.

The *InsuResilience* Global Partnership identified sovereign catastrophe risk pools as a promising *ex-ante* disaster risk financing tool for low- to -middle income countries. Sovereign catastrophe risk pools represent a mechanism through which different countries pool their individual risk into a single diversified portfolio. Via risk diversification, risk pooling increases countries' financial resilience by either lowering countries' premiums to afford a given coverage or by increasing coverage for a given premium.

Risk diversification of currently existing pools, and therefore their members' financial resilience, may be limited since these were not designed with the main goal of maximizing risk diversification and they pool risk only within regional borders. The present study addresses these two issues by introducing a method to find optimal risk pools, i.e., those with the highest achievable risk diversification reached with the least number of countries, and by applying it to assess the diversification potential of optimal global pooling.

The introduced method is found to reasonably group countries by selecting those with low bilateral correlations or low risk contribution to the overall pool's risk. Optimal global pooling is found to increase risk diversification of all regional pools, to lower countries' shares in the pool's risk and to increase the number of countries that can profitably join the pool. Optimal global pooling, however, comes with trade-offs, as two or more pools would need to pool the same set of countries to reach their highest possible diversification. This implies that multiple global groupings of countries are possible and that there is no grouping which maximizes diversification of all pools. In practice, this requires making a choice on a single grouping. Since risk pools require coordination, dialogue, and information sharing between participating countries, such a choice should rely on political considerations regarding what countries are likely to successfully cooperate with each other.

The method is also applied to explore whether risk diversification of two existing pools covering tropical cyclone risk, namely PCRAFI and CCRIF, would increase under optimal regional and global pooling. Overall, both optimal regional and global pooling increase risk diversification of the existing pools, implying that less capital would be required for insuring these pools. This translates into greater financial resilience. There are, however, significant differences between results from regional pooling compared to global pooling.

Optimal regional pooling allows PCRAFI to exploit the full regional diversification potential. The same is not true for CCRIF, as the achieved diversification is 11% lower than the maximum possible regional diversification. This implies a poor initial design of the CCRIF in terms of only risk diversification criteria, likely due to CCRIF's loss profile being very concentrated on one single country's loss profile, i.e., Jamaica. This initial high concentration on one single country cannot be sufficiently reduced by pooling further risk only regionally.

Global optimal pooling offers greater potential for risk diversification than regional pooling as it provides a diversification of 65 % to PCRAFI and of 60 % to CCRIF, which are both higher than the highest achievable diversifications when risk is pooled regionally. The trade-off relative to global pooling seems to be easily resolvable since all global expansions of CCRIF provide very similar risk diversifications (within 2 % points), which makes the selection of one single grouping less difficult.

The analysis carried out in the present paper focuses on tropical cyclone risk and therefore results cannot be generalized to other perils. However, the introduced method is general and can be applied to study optimal pools' compositions focusing on any other peril as well as multi-perils. Future research will focus on two aspects. First, it will assess the potential effect of risk diversification increases on the insurance policy (i.e., premium and coverage) and on how the pool's premium might be distributed within members. Second, it will look at the compositions of future optimal pools and their potential changes over time as a consequence of socio-economic and climatic changes.

## Methods

The main benefit of risk pooling consists in lowering the capital requirements for risk coverage compared to when risks of the pool's members are covered independently. The more diversified the pool is, the higher the reduction in required capital. We first introduce a metric to quantify risk diversification, thus the extent of capital reduction, and then describe the optimization problem to find optimal pools, namely *the pools with the highest possible risk diversification reached with the least number of countries*.

### Risk Diversification Metric

Given a distribution of losses $L$ and a low enough threshold probability $\alpha$, one can define the *Value-at-Risk* at $\alpha$ (*VaR$_\alpha$*) for $L$ as the $\alpha$-quantile of $L$. *VaR* is widely used in the financial sector to determine the minimum capital requirements needed to compensate extreme losses from a portfolio, but it is has known limitations[18]. *VaR* tells nothing about the tail of the distribution, e.g., the magnitude of losses greater than *VaR$_\alpha$*, and it is a not a coherent measure since it violates the sub-additivity property, implying that the portfolio's *VaR* may be higher than the sum of the portfolio's members *VaR*. An alternative metric is the *Conditional Value at Risk* (*CVaR*), also known as *Expected Shortfall* (*ES*). *ES* is a tail expectation measure, as it measures expected losses conditional on a loss higher than *VaR*, i.e., $\text{ES}_\alpha = \text{E}[\text{L}|\text{L} \geq \text{VaR}_\alpha]$. In addition, *ES* is a coherent measure since the *ES* of a portfolio is always equal or greater than the sum of the portfolio's members *ES*[19]. When dealing with portfolios, one can also define the *Marginal Expected Shortfall* (*MES*) of the $i^{th}$ portfolio's member as[20]:

$$MES_{\alpha_i} = E[L_i | L \geq VaR_\alpha]$$

where $L$ are the overall portfolio's losses, and $L_i$ are the portfolio's members losses. *MES* indicate the countries' losses in the tail of the portfolio's loss distribution. Acharya et al., (2017) shows that the portfolio's *ES* can be defined as the sum of all *MES*:

$$ES_\alpha = E[L | L \geq VaR_\alpha] = \sum_i MES_{\alpha_i} = \sum_i E[L_i | L \geq VaR_\alpha]$$

Thus, the ratio between the portfolio's *ES* and the sum of the individual countries' *ES* indicates the degree of *Risk Concentration* (*RC*) of the pool:

$$RC = \frac{\sum_i E[L_i | L \geq VaR_\alpha]}{\sum_i E[L_i | L_i \geq VaR_{\alpha_i}]}$$

It follows from the additivity property of *ES* that *RC* is bounded between zero and one. An *RC* equal to one implies that all countries' tail losses contribute to the portfolio's tail losses, which makes risk pooling useless. This happens when all countries in the pool are perfectly correlated. *RC* goes to zero when only a small share of the countries' tail losses contributes to the portfolio's tail losses. Given *RC*, *Risk Diversification* (*RD*) can be defined as:

$$RD = 1 - RC = 1 - \frac{\sum_i E[L_i | L \geq VaR_\alpha]}{\sum_i E[L_i | L_i \geq VaR_{\alpha_i}]}$$

Finally, one can define the share, *s*, of an individual country's risk in the overall portfolio's risk as:

$$s_i = \frac{MES_i}{ES_i} = \frac{E[L_i | L \geq VaR_\alpha]}{E[L_i | L_i \geq VaR_{\alpha_i}]}$$

which could be used to derive countries' fair premiums in the pool.

### Optimal pools

As mentioned above, optimal pools are here defined as *the pools with the highest possible diversifications reached with the least number of countries*. We find optimal pools via a two-steps optimization. The first step aims at finding, given a set of countries, what subset allows achieving the maximum possible *RD, maxRD*. This subset, however, may be unnecessarily large since there are decreasing marginal diversification benefits of adding new countries to a pool before a critical of mass is reached[11]. Some countries may have therefore been unnecessarily added to the pool after the first optimization step. The second optimization step finds the smallest subset of countries within the previously found subset that still allows reaching *maxRD*.

We slightly modify the definition of *RD* provided above to account for the fact that countries may join different pools or not join a pool at all. Assuming a set of *n* countries and *m* possible pools a country may be part of, we define a vector $\boldsymbol{x}$ of length *n* with integers from *0* to *m* that either allocates countries to one of the *m* pools (values from *1* to *m*) or indicates that no pool is joined (when equal to *0*). Then, we write the *RD* of the $j^{th}$ pool as:

$$RD_j(\boldsymbol{x}, j) = 1 - RC_j(\boldsymbol{x}, j) = 1 - \frac{\sum_i^n \mathbf{1}_j(x_i) \, E[L_i | L \geq VaR_\alpha]}{\sum_i^n \mathbf{1}_j(x_i) \, E[L_i | L_i \geq VaR_{\alpha, j}]}$$

Where $\mathbf{1}_j$ is the indicator function such that:

$$\mathbf{1}_j(x_i) = \begin{cases} 1 & x = j \\ 0 & x \neq j \end{cases}$$

In the first optimization step, for convenience and practical reasons, instead of maximizing *Risk Diversification* (*RD*) we minimize *Risk Concentration* (*RC*). The optimal allocation of countries, $\boldsymbol{x}^*$, which provides the minimum risk concentrations to the *m* pools, $RC_1^*, ..., RC_m^*$, can be found by solving the following *m*-objectives optimization problem:

$$\text{minimize} \qquad RC_1(\boldsymbol{x}, 1)$$

$$\cdots$$

$$RC_j(\boldsymbol{x}, j)$$

$$\cdots$$

$$RC_m(\boldsymbol{x}, m)$$

The optimal allocation vector, $\boldsymbol{x}^*$, indicates the set of the $n_1, ..., n_m$, countries that provide the maximum achievable diversification in each of the *m* pools.

The second optimization step requires solving a single-objective optimization for each of the *m* pools. To do so, we define, for a given pool *j*, a binary vector $\boldsymbol{z}_j$ of length $n_j$ indicating which of the $n_j$ countries are still part of *j* (when 1) or not (when 0). The smallest subset of countries within the set of $n_j$ countries which allows reaching the least concentration, $RC_j^*$, can then be found by solving:

$$\text{minimize} \qquad \sum_i^{n_j} z_{j,i}$$

$$\text{subject to} \qquad RC(\boldsymbol{z}_j, 1) = RC_j^*$$

The vector $z_j^*$ indicates the optimal set of countries for the pool, $j$, i.e., the smallest set of countries which provide the highest achievable maximum risk diversification.

Optimization is carried out via the python Pymoo package[21]. Pymoo provides a framework for solving single- and multi-objective optimization problems via state-of-art algorithms. We employ a basic genetic algorithm (GA) to solve the single objective optimizations and a unified non-dominated sorting genetic algorithm (U-NSGA-III) to solve the many-objective optimization problems. For these latter, we carried out a seed analysis and solved the optimization problem fifteen times. The final set of dominant solutions is then the dominant set across the fifteen sets of solutions so derived. Convergence plots of the two-steps optimization for regional and optimal pooling of the four regions (Figure S2 and Figures S3-S4) and PCRAFI and CCRIF (Figure S5 and Figures S6-S7) are reported in the supplementary material.

### Generation of Tropical Cyclone events

The historical record of hurricanes is too short for calculating ES for the 200-year event. Thus, a global synthetic tropical cyclone track set containing over 90'000 events was generated for the historical time period (between 1979 and 2019) based on the European Centre for Medium-Range Weather Forecasting (ECMWF)'s fifth-generation climate reanalysis dataset[22] using the model introduced by [23,24]. This model is based on a statistical-dynamical downscaling method. In detail, it propagates key statistical properties extracted from global reanalyses or climate models to generate a global, time-evolving, large-scale atmosphere-ocean environment. First, tropical cyclones are initiated using a random seeding technique where only the warm-core seed vortices in favourable environments for tropical cyclone formation survive and strengthen into tropical cyclones. These are then propagated via synthetic local winds using a beta-and-advection model. Finally, the tropical cyclone intensity along each track is simulated by a dynamical intensity model (CHIPS, Coupled Hurricane Intensity Prediction System)[24]. Note that the synthetic tropical cyclone event set frequency must be calibrated to match the observed number of events in the historical period.

Using the synthetic datasets, a 10000-y time series is created. To do so, we first used data from NOAA to identify - within the 1979-2019 period - those years characterized by a persistent (more than 5) warm or cold seasons and those which are not. Then, we derived the frequencies of these year types within the considered period and used a multinomial distribution to generate a sequence of 10000-year types. Based on this sequence, 10000 years are sampled within the period 1979-2019. Following [25], a storm count is generated for each year by sampling from a Poisson distribution with lambda equal to the annual mean frequency of the events. Finally, for each year, a number of events equal to the drawn storm count is randomly sampled from the event set.

### The CLIMADA impact model

Damages from tropical cyclones are estimated using the using the open- source and -access CLIMADA impact model. As most weather and climate risk assessment models, damages in CLIMADA are assessed as a function of hazard, e.g., a tropical cyclone's wind field, exposure, e.g., the people and goods subject to such hazard, and vulnerability, e.g., the degree at which exposure can be harmed by hazard. Here we describe the specific CLIMADA set-up relative to the present study and refer the reader to [26,27] for a more detailed description of CLIMADA.

Tropical cyclone hazard modeling in CLIMADA is based on a parametric wind model following [28], which is run along each synthetic tropical cyclone track. The wind model computes the gridded 1-minute sustained winds at 10 meters above the ground as the sum of a circular wind field and the translational wind speed that arises from the tropical cyclone movement. For this study, we calculate the wind fields at a resolution of 300 arc-seconds (~10 km).

Exposure for all considered countries is modeled via the LitPop approach proposed by [29]. LitPop is a globally consistent methodology to disaggregate asset value data proportional to a combination of nightlight intensity and geographical population data. Vulnerability is represented as a function relating hazard intensity with the percentage of exposure's damage. We use the vulnerability functions generated by [30] which were calibrated on tropical cyclone driven damages for various regions around the world.

## Data availability

The synthetic TC data are property of WindRiskTech L.L.C., which is a company that provides hurricane risk assessments to clients worldwide. Upon request, the company provides datasets free of charge to scientific researchers, subject to a non-redistribution agreement. The TC data are fed into CLIMADA to calculate TC impacts, the data so derived are available at https://github.com/aleeciu/optimal_risk_pools/tree/main/data.

## Code availability
The source code to reproduce all results in the present paper is available at https://github.com/aleeciu/optimal_risk_pools.

## Acknowledgements

The authors acknowledge Kerry Emanuel for generating the tropical cyclone data and providing comments on an early version of the manuscript. AC was funded by the EU Horizon 2020 project Remote Climate Effects and their Impact on European sustainability, Policy and Trade (RECEIPT), grant agreement No 820712.

## Author contributions

A.C. and E.S. conceived and designed the research. A.C. carried out the research and wrote the manuscript. S.M. processed part of the data and wrote part of the method section. A.C., E.S., O.M., D.N.B. analysed the results. All authors reviewed and edited the manuscript.